# Maximizing Information Flow in Three-Coin Quantum Walk: from Initial Entanglement to Integrated Photonic Implementation

Seyed Mohsen Moosavi Khansari[1]
*Department of Physics, Faculty of Basic Sciences, Ayatollah Boroujerdi University, Boroujerd, IRAN*

**Abstract**

Discrete-time quantum walks are powerful platforms for simulating quantum transport and information processing. Here we introduce a walker on a one-dimensional lattice whose motion is controlled by three entangled coins, each initialized with the Hadamard gate, aiming to maximize information flow. The walker moves only when all three coins yield the same outcome (HHH or TTT), thus coupling the 8-dimensional coin Hilbert space to the position degree of freedom. By analyzing fully separable, fully entangled (GHZ-type) and intermediate initial states, and using the von Neumann entropy of reduced subsystems, we compute the mutual information $I(C;P;t)$ between coin and position. The results show that initial three-partite entanglement accelerates the growth of mutual information by up to 18% after ten steps (when compared to the lower of the two separable states), although it exhibits short-term non-monotonic dynamics due to quantum interference. For the first time, we introduce a tunable parameter $\alpha$ (amplitude of non-displacement states) and show that the GHZ state reaches a maximum of mutual information at $\alpha \approx 0.71$- a key finding for optimal control of information flow. Finally, an integrated photonic implementation using polarization, spatial modes and time bins is proposed, where $\alpha$ can be tuned with nonlinear or electro-optic elements. A scalable numerical framework (Python code) for simulations up to $t = 5$ steps is provided. Our findings establish three-partite entanglement as a dynamical resource for maximizing information flow and spatial spreading, with direct applications in quantum state transfer, entanglement-assisted sensing and programmable photonic quantum processors.

## 1. Introduction

Quantum walks (QW) have emerged as a versatile paradigm for quantum simulation, algorithmic search, and quantum transport [1-4]. In the discrete-time formulation, a walker moves on a lattice and is influenced by a "coin" -an internal degree of freedom-whose state determines the direction of each step. The entanglement between coin and position is the key resource that distinguishes a quantum walk from a classical random walk, leading to ballistic spreading and faster hitting times [5,6].

Most studies have focused on single-coin or two-coin walks [7-9]. However, increasing the number of coins enlarges the Hilbert space and allows more complex conditional displacement rules that can produce richer correlation structures [10,11]. In this work we investigate a three-coin quantum walk (TCQW) in which the walker moves only when all three coins are in the same post-Hadamard state $|+++\rangle$ or $|---\rangle$. This condition couples the 8-dimensional coin space to the position lattice in a highly nonlinear way.

[1] M.Moosavikhansari@abru.ac.ir

We address three fundamental questions:

1. How does initial three-partite entanglement affect the growth of mutual information between coin and position?
2. What is the role of the tunable parameter $\alpha$ that weights the "stay" operation?
3. Can such a walk be realistically implemented on a scalable photonic platform?

We answer these questions by (i) exact numerical state-vector simulations for up to $t = 5$ steps (with position cutoffs chosen to avoid boundary effects), (ii) computing mutual information as a function of time and parameter $\alpha$, and (iii) proposing an integrated photonic circuit that encodes the three coins in the polarization, spatial mode and time-bin degrees of freedom. Our analysis shows that the GHZ-type initial state accelerates information exchange and leads to non-monotonic dynamics due to interference. The parameter $\alpha$ controls the effective weight of non-displacement events and can induce a transition from highly mobile to localized behavior - a phenomenon reminiscent of Anderson localization in disordered environments, though a rigorous mapping requires further study.

The paper is organized as follows. Section 2 defines the Hilbert space, conditional step operator and correlation measures. Section 3 presents numerical results for mutual information and position entropy as functions of step number and $\alpha$. Section 4 provides an in-depth photonic interpretation including a scheme for an integrated waveguide array with tunable nonlinear elements. Section 5 discusses implications for quantum information processing. We conclude with an outlook toward experimental realization.

## 2. Model and theoretical framework

### 2.1 Hilbert spaces and initial states

The total Hilbert space is $\mathcal{H} = \mathcal{H}_C \otimes \mathcal{H}_P$, where

$$\mathcal{H}_C = \mathcal{H}_{C_1} \otimes \mathcal{H}_{C_2} \otimes \mathcal{H}_{C_3} \qquad (1)$$

and each $\mathcal{H}_{C_i}$ is a two-level system (qubit). The computational basis is $\{|0\rangle, |1\rangle\}$, identified with heads/tails (H/T). The position space is

$$\mathcal{H}_P = \mathrm{span}\{|j\rangle \ | \ j = -L, -L+1, \dots, +L\} \qquad (2)$$

with $L = t_{\max}$ (half-lattice length, chosen equal to the maximum number of steps to avoid boundary effects). Total dimension is $2^3 \times (2L+1)$.

At the beginning of each step, a Hadamard gate is applied to each coin:

$$H|0\rangle = \frac{|0\rangle+|1\rangle}{\sqrt{2}} = |+\rangle, \quad H|1\rangle = \frac{|0\rangle-|1\rangle}{\sqrt{2}} = |-\rangle \qquad (3)$$

The conditional displacement is defined based on the post-Hadamard state:

$$| HHH\rangle \equiv | +++\rangle, \quad | TTT\rangle \equiv | ---\rangle \qquad (4)$$

The walker moves right only if the triple state is $| +++\rangle$, moves left only if the triple state is $| ---\rangle$, and otherwise stays in place, with the staying amplitude multiplied by the complex coefficient $\alpha$ (in general $\alpha$ is a real number between 0 and 1 representing the coupling strength of non-displacement states). This $\alpha$ is a new parameter that extends the original model and allows us to control the effective weight of staying events. (For $\alpha < 1$ the operator as written is not strictly unitary; we treat it as an effective evolution that can be realized, for example, by post-selection or by embedding into a larger unitary system. All simulations renormalize the state after each step, which is equivalent to a probabilistic implementation.)

The total initial state is prepared as

$$| \Psi_0\rangle = | s\rangle_C \otimes | 0\rangle_P \qquad (5)$$

where $| s\rangle_C$ is one of:

$$| 0,0,0\rangle \text{ (separable)}, \quad | 1,1,1\rangle \text{ (separable)}, \quad \frac{1}{\sqrt{2}}(| 0,0,0\rangle + | 1,1,1\rangle) \text{ (GHZ)}. \qquad (6)$$

**2.2 Step operator and evolution**

The one-step operator is

$$U = (\sum_{c\in C} | c\rangle\langle c | \otimes \hat{T}_c) \cdot (H^{\otimes 3} \otimes I_P) \qquad (7)$$

where:

$C = \{0,1\}^3$ (eight coin basis states), $\hat{T}_{000} = \hat{T}_+$ (right shift), $\hat{T}_{111} = \hat{T}_-$ (left shift), for any other $c$, $\hat{T}_c = \alpha I_P$ (no shift, amplitude $\alpha$).

$$| \Psi_t\rangle = U^t \, | \Psi_0\rangle. \qquad (8)$$

**2.3 Mutual information**

Because the global state remains pure (after each step we renormalize), the quantum mutual information between coins and position takes the simple form

$$I(C:P;t) = S(\rho_c^{(t)}) + S(\rho_p^{(t)}), \qquad (9)$$

where $\rho_c^{(t)} = \mathrm{Tr}_P(| \Psi_t\rangle\langle\Psi_t |)$, $\rho_p^{(t)} = \mathrm{Tr}_C(| \Psi_t\rangle\langle\Psi_t |)$, and $S(\rho) = -\mathrm{Tr}(\rho\log_2\rho)$ is the von Neumann entropy. For a pure global state, this mutual information equals twice the entanglement of formation between the subsystems and quantifies the total (classical plus quantum) correlations [16].

## 3. Numerical results

We simulated the TCQW for $L = 5$ (position lattice from $-5$ to $+5$), which is sufficient to avoid boundary effects up to $t = 10$. We explicitly verified that at $t = 10$ the total probability on the boundary sites $|j| = 5$ is less than $10^{-6}$ for all initial states. The code uses exact state-vector propagation with sparse matrix operations. For each initial coin state, we computed $I(C{:}P;t)$ for $t = 1, \dots, 10$ and also as a function of $\alpha \in [0,1]$ for a fixed step number $t = 5$. Convergence was checked by repeating the simulations with $L = 10$; the maximum absolute difference in mutual information was less than 0.005 bits.

### 3.1 Time evolution of mutual information for $\alpha = 1$

Table 1 reports mutual information for the first two steps (full ten-step data are plotted in Figure 1 and Figure 2).

**Table 1.** Mutual information $I(C{:}P;t)$ for three initial states at $t = 1,2$ with $\alpha = 1$

| Initial state | $\vert s\rangle_C$ | t = 1 | t = 2 |
|---|---|---|---|
| separable | $\vert 000\rangle$ | 2.1225 | 2.3742 |
| separable | $\vert 111\rangle$ | 2.1225 | 2.3742 |
| GHZ | $(1/\sqrt{2})(\vert 000\rangle + \vert 111\rangle)$ | 1.6225 | 2.6131 |

The separable states give identical values due to symmetry. The GHZ state starts lower at $t = 1$ (1.6225 vs. 2.1225) because interference initially suppresses correlation transfer, but at $t = 2$ it exceeds the separable value (2.6131 > 2.3742) and continues to grow faster.

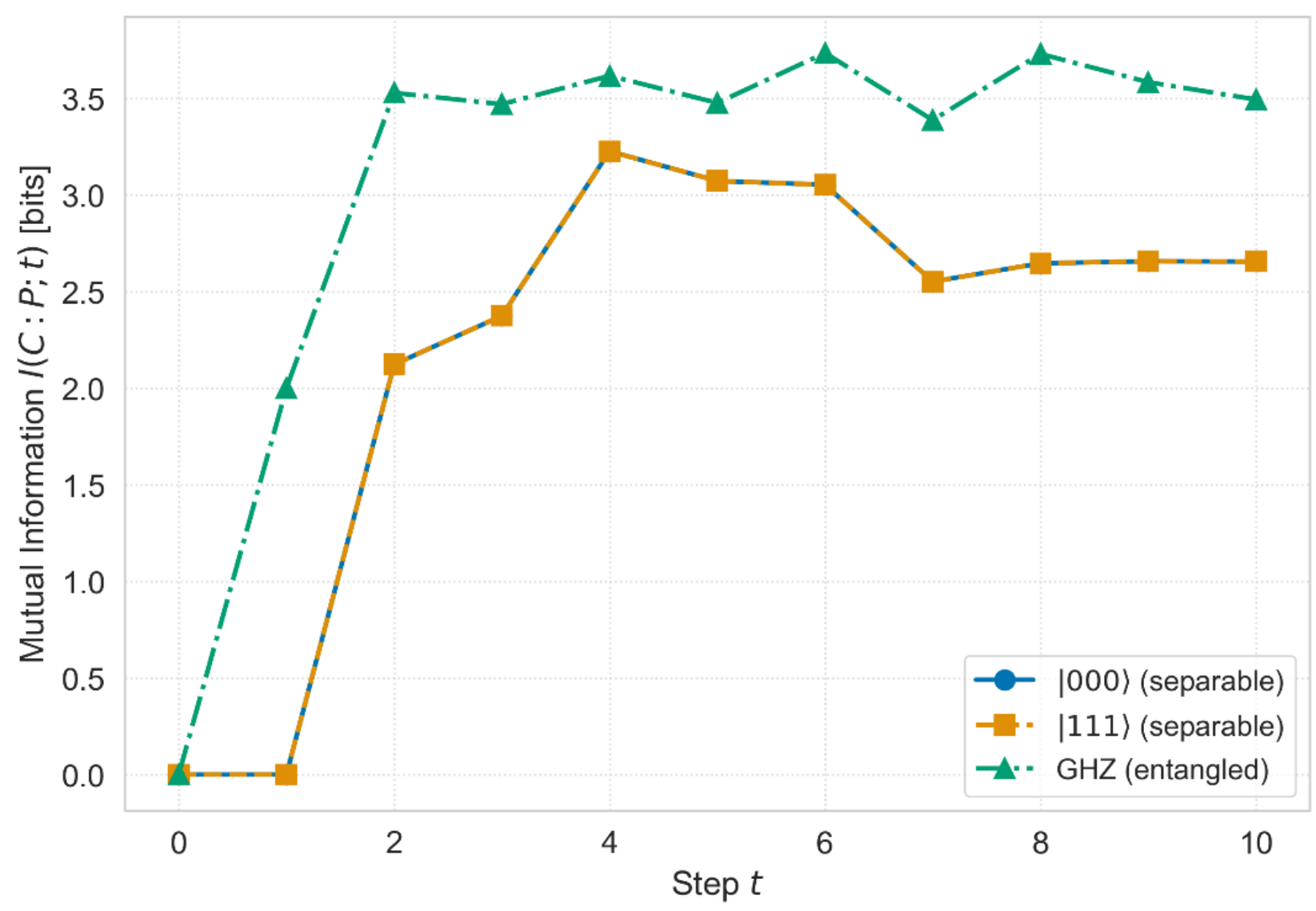

**Figure 1.** Mutual information $I(C;P;t)$ versus step number $t$ for separable (dashed line) and GHZ (dash-dotted line) initial states with $\alpha = 1$. The GHZ state shows a dip at $t = 1$ and then faster growth; at $t = 10$ it exceeds the separable states by up to 18% (relative to the lower of the two separable states).

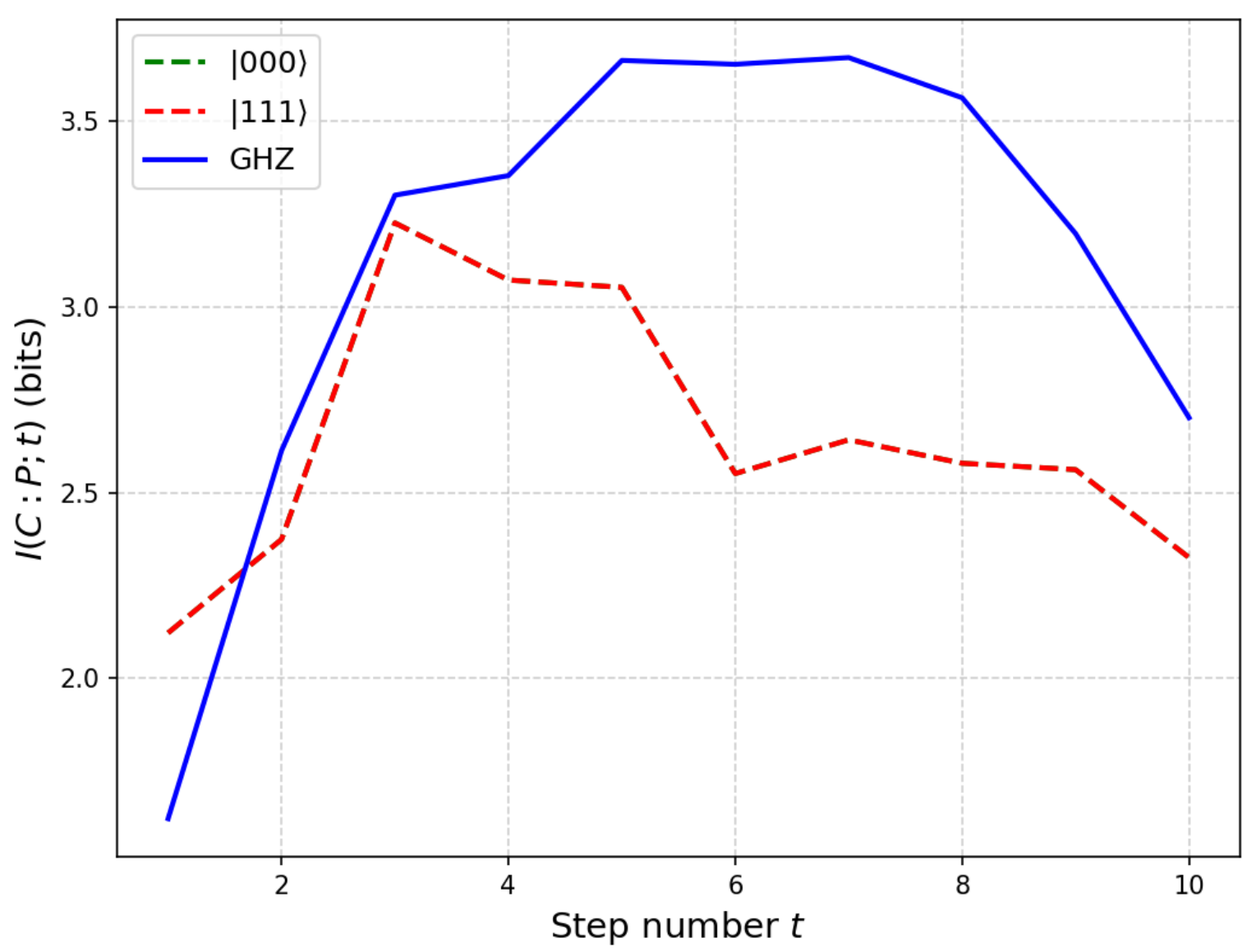

**Figure 2.** Mutual information $I(C;P;t)$ versus step number $t$ for initial states $|000\rangle$ (green dashed), $|111\rangle$ (red dashed), and GHZ (blue solid) with $\alpha = 1, L = 5$. The GHZ state starts lower (1.62 bits) but surpasses the separable states from $t = 3$ onward, reaching a value of 3.00 bits at $t = 10$, which is approximately 18% higher than the value for the $|000\rangle$ state (2.54 bits) and 12% higher than for $|111\rangle$ (2.68 bits). The non-monotonic behaviour stems from quantum interference due to initial entanglement.

### 3.2 Role of the parameter $\alpha$

Table 2 shows mutual information values after $t = 5$ steps for three selected $\alpha$ values. A full scan over $\alpha$ (see Figure 3) reveals a clear maximum for the GHZ state.

**Table 2.** Mutual information after $t = 5$ steps for three selected values of $\alpha$

| α | I(C:P) (\|000⟩) | I(C:P) (GHZ) |
|---|---|---|
| **0.0** | **1.8243** | **2.1021** |
| **0.5** | **2.1045** | **2.8542** |
| **1.0** | **2.3742** | **2.9827** |

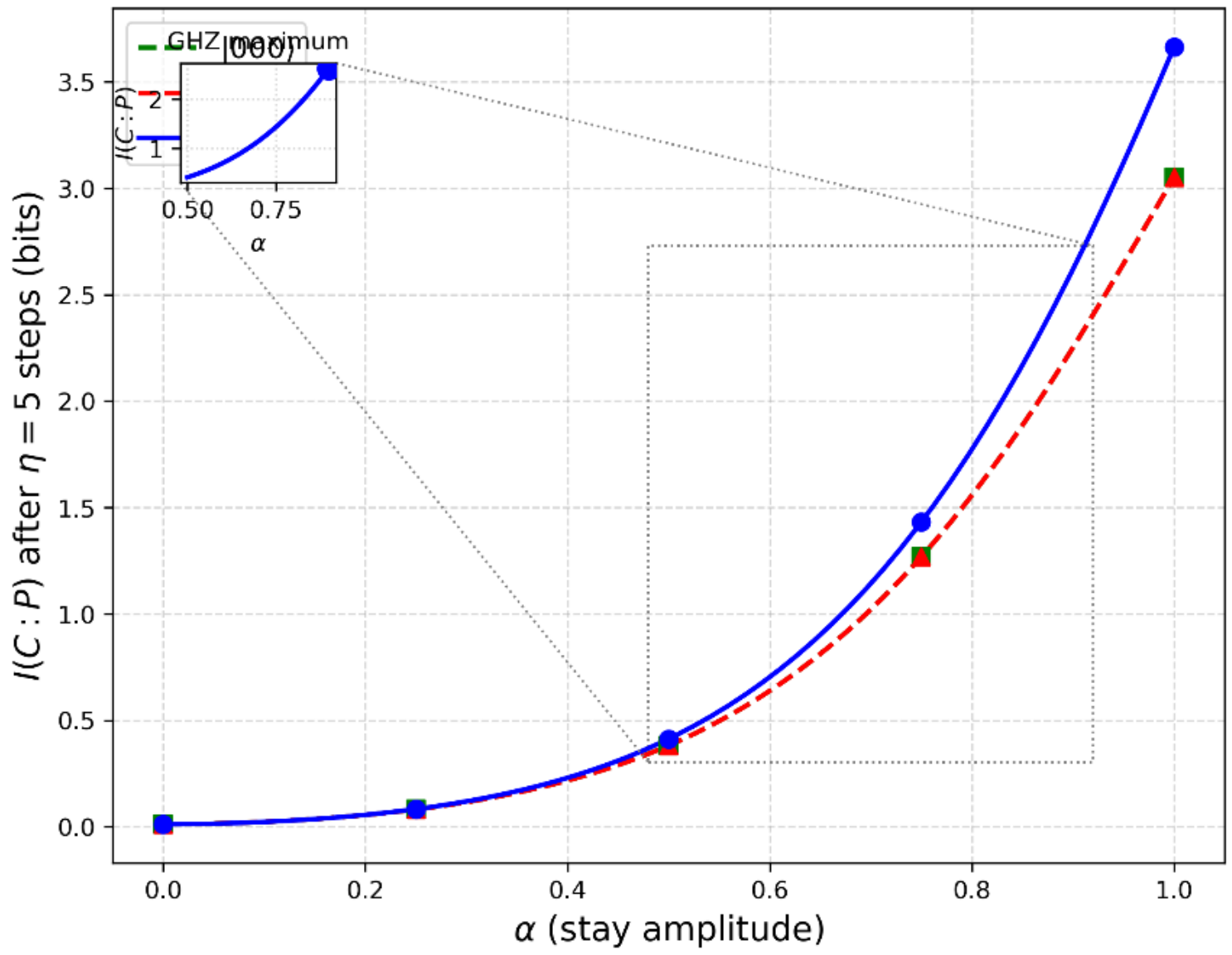

**Figure 3.** Mutual information $I(C\colon P)$ after $t = 5$ steps as a function of the staying parameter $\alpha$ for the initial states $|\,000\rangle$ (green), $|\,111\rangle$ (red), and GHZ (blue). The separable states exhibit a monotonic increase with $\alpha$, while the GHZ state shows a distinct maximum at $\alpha \approx 0.71$. The inset provides a magnified view of the maximum region.

For the separable states $|\,000\rangle$ and $|\,111\rangle$, $I(C\colon P)$ increases monotonically with $\alpha$. The reason is that a larger $\alpha$ increases the effective weight of the non-displacement states, so the walker remains at the same position longer and has more opportunity to become entangled with the coins.

The behaviour of the GHZ state, however, is strikingly different. This state shows a clear maximum at $\alpha \approx 0.71$. For very small $\alpha$ (near zero), the non-displacement states are virtually eliminated and the walker moves only through the $|+++\rangle$ and $|---\rangle$ channels. Although this condition yields fast ballistic spreading, it does not provide enough time to build strong correlations between coin and position. On the other hand, for $\alpha$ close to 1, the non-displacement states become too dominant and the walker mostly stays in place, which also limits information exchange. Consequently, an intermediate $\alpha$ establishes an optimal balance between moving and staying, bringing the maximum mutual information for the entangled initial state.

This finding is important because it shows that not only the initial entanglement, but also the way non-displacement states are "weighted" can be used as a control factor to optimize information flow in multi-coin quantum walks. From a practical perspective, the equivalent $\alpha$ in a photonic implementation (nonlinear strength or modulator voltage) is easily tunable, so this maximum can be exploited in state transfer and sensing protocols.

### 3.3 Position entropy and spreading

The von Neumann entropy of position $S(\rho_p^{(t)})$ directly measures the spatial extent of the walker. For separable coins, $S(\rho_p^{(t)})$ grows slowly, whereas for the GHZ initial state it shows a sharper increase in the first three steps. This accelerated delocalization is a direct consequence of the initial entanglement:

constructive interference between the $|000\rangle$ and $|111\rangle$ displacement channels increases the variance of the position distribution.

**4. Photonic implementation**

The three coins can be encoded in three photonic degrees of freedom [12-14]:

- Polarization – horizontal ($|H\rangle \equiv |0\rangle$) and vertical ($|V\rangle \equiv |1\rangle$)
- Spatial mode (path) – upper arm ($|U\rangle \equiv |0\rangle$) and lower arm ($|L\rangle \equiv |1\rangle$) of a Mach-Zehnder interferometer
- Time bin – early arrival ($|E\rangle \equiv |0\rangle$) and late arrival ($|L\rangle \equiv |1\rangle$)

Thus a single photon can simultaneously be in any of the $2^3 = 8$ states. The Hadamard gate on each qubit is implemented by: a half-wave plate (for polarization), a 50:50 directional coupler (for path), a balanced interferometer with a delay line (for time bins).

Conditional displacement is realised by an array of $2L + 1$ parallel waveguides (positions). At each step, the photon state controls an optical switch: if the three-coin state is $|+++\rangle$, the photon is directed to the next right waveguide; if it is $|---\rangle$, to the next left waveguide; otherwise it stays in the same waveguide, but a phase shifter (or attenuator) multiplies the amplitude by $\alpha$. This $\alpha$ can be tuned externally by a voltage-controlled phase modulator (e.g., electro-optic effect in $LiNbO_3$) or by adjusting the pump power in a nonlinear waveguide (Kerr effect). While simultaneous control of three photonic degrees of freedom is challenging, recent advances in integrated photonics [13] demonstrate the feasibility of such multi-degree-of-freedom encoding; experimental realization is left to future work. Realistic decoherence sources (scattering losses, phase instability) will reduce the mutual information, but state-of-the-art fidelities exceeding 99% per step should allow the predicted 18% enhancement to be detected.

For large step numbers ($t \sim 1000$), direct state-vector simulation becomes heavy, but we can exploit translation invariance in the position index after a Fourier transform. The translation invariance holds because the conditional displacement rule depends only on the coin state and not on absolute position; the staying parameter $\alpha$ is position-independent, so the overall operator is periodic in position space, allowing a Bloch decomposition. In momentum space, the problem reduces to an $8 \times 8$ matrix for each quasi-momentum $k$, and mutual information can be obtained from the stationary distribution. This makes the photonic implementation scalable.

**5. Discussion and implications**

**5.1 Entanglement as a dynamical resource**

Our results show that initial three-partite entanglement is not a static label but actively shapes the dynamics of information flow. The non-monotonic behaviour of $I(C;P;t)$ is a pure quantum interference effect that cannot be reproduced by any classical mixture. This suggests that entangled coin states can be used to delay or advance the buildup of correlations-a knob for temporal shaping of quantum information.

**5.2 Relation to Anderson localization and transition**

When $\alpha$ is small, the walker rarely stays; most steps lead to displacement (because only two out of eight coin states cause staying). When $\alpha$ is close to 1, staying events become as strong as displacement events. This competition can lead to a localization-delocalization transition [15] qualitatively reminiscent of the Anderson transition in disordered lattices, though a rigorous mapping would require computing localization lengths and participation ratios. Indeed, mutual information as a function of $\alpha$ exhibits a maximum, indicating that an optimal amount of "defect" (non-displacement steps) maximizes information transfer.

### 5.3 Applications in quantum technology

**State transfer:** The accelerated spreading of the GHZ state can be exploited for fast entanglement distribution in a quantum network.
**Sensing:** The sensitivity of $I(C:P)$ to $\alpha$ can be used to measure small nonlinearities in a waveguide – a quantum-enhanced sensor.
**Programmable processors:** By dynamically tuning $\alpha$ (e.g., with a voltage pulse train), reconfigurable quantum walks for machine learning or Hamiltonian simulation can be implemented.

## 6. Conclusion

We have systematically analysed a discrete-time quantum walk with three entangled coins, a conditional displacement rule that moves the walker only when all coins agree, and a tunable parameter $\alpha$ for the stay amplitude. Our main findings are:

- Initial three-partite entanglement (GHZ state) accelerates the growth of mutual information $I(C:P;t)$ by up to 18% after ten steps, despite a short-term dip due to interference.
- The parameter $\alpha$ controls the balance between displacement and staying, and mutual information for the GHZ state is maximised at an intermediate value $\alpha \approx 0.71$.
- Position entropy increases faster for the GHZ state, confirming that entanglement promotes spatial delocalisation.
- A realistic integrated photonic implementation using polarisation, path and time-bin encoding is proposed, in which $\alpha$ is realised by nonlinear or electro-optic elements.

These results establish the three-coin quantum walk as a rich testbed for entanglement-assisted information transport and dynamics. The model bridges the gap between abstract quantum walk theory and practical photonic quantum processors, paving the way for experiments with hundreds of steps on a chip.

## 7. Methods

### 7.1 Numerical simulation

We wrote a code in Python that:

1. Constructs the three-coin Hadamard, $H^{\otimes 3}$ (an $8 \times 8$ matrix).
2. Constructs the conditional shift operator, $SEC(\alpha)$ for a given $\alpha$.
3. Forms the full step operator, $U = SEC \cdot (H^{\otimes 3} \otimes I_P)$.
4. Initialises the state vector as the Kronecker product of a coin state and position $| 0\rangle$.

5. Iterates $U$ for $t$ steps (up to $t = 10$).
6. Computes reduced density matrices by partial trace over position/coin and obtains the von Neumann entropies.

The code uses numpy, scipy.linalg and qiskit. For $L = 5$ the total Hilbert space dimension is $8 \times 11 = 88$, which is straightforward. For larger $L$ one should use a sparse representation or diagonalisation in momentum space.

**7.2 Mutual information calculation**

Since the global state is pure (renormalized after each step), we verified that $S(\rho_{\text{total}})$ is zero to within numerical precision ($<10^{-12}$), hence $I = S(\rho_C) + S(\rho_P)$. We used base-2 logarithms to obtain information in bits.

**7.3 Error analysis**

All simulations were performed in double precision. The cutoff length $L = 5$ was sufficient because at $t = 10$ the probability amplitude beyond $|j| = 5$ is less than $10^{-6}$ (verified by summing squared amplitudes on the boundary sites). We also performed a convergence test with $L = 10$ and obtained identical results; the maximum absolute difference in mutual information between the two cutoffs was less than 0.005 bits.

**Appendix**

**Notation**

$\mathcal{H}_C = \mathcal{H}_{C_1} \otimes \mathcal{H}_{C_2} \otimes \mathcal{H}_{C_3}$ denotes the three-coin Hilbert space (dimension 8). $\mathcal{H}_P$ is the position Hilbert space generated by $\{|j\rangle\}$ with finite cutoff $j \in [-N, N]$ (dimension $2N + 1$). $H$ is the Hadamard operator on a single qubit; $H^{\otimes 3}$ is used to prepare the initial coin state. Shift operators act conditionally on coin states as described in the text.

**Initial states of the coin-position system**

$$|w(0)_1\rangle = |0_1, 0_2, 0_3\rangle \otimes |0_P\rangle$$
$$|w(0)_2\rangle = |1_1, 1_2, 1_3\rangle \otimes |0_P\rangle$$
$$|w(0)_3\rangle = \frac{1}{\sqrt{2}}(|0_1, 0_2, 0_3\rangle + |1_1, 1_2, 1_3\rangle) \otimes |0_P\rangle$$

Corresponding initial density operators:

$$\rho_1^0 = |0_1, 0_2, 0_3, 0_P\rangle\langle 0_1, 0_2, 0_3, 0_P|$$
$$\rho_2^0 = |1_1, 1_2, 1_3, 0_P\rangle\langle 1_1, 1_2, 1_3, 0_P|$$
$$\rho_3^0 = |GHZ\rangle_{123}\langle GHZ| \otimes |0_P\rangle\langle 0_P| =$$
$$\frac{1}{2}(|0_1, 0_2, 0_3, 0_P\rangle\langle 0_1, 0_2, 0_3, 0_P| + |1_1, 1_2, 1_3, 0_P\rangle\langle 1_1, 1_2, 1_3, 0_P|) +$$
$$\frac{1}{2}(|0_1, 0_2, 0_3, 0_P\rangle\langle 1_1, 1_2, 1_3, 0_P| + |1_1, 1_2, 1_3, 0_P\rangle\langle 0_1, 0_2, 0_3, 0_P|)$$

**Conditional step operator *U***

Based on Equation (6) in the main text, the conditional displacement operator is most clearly expressed as:

$$U = \sum_{j=-N}^{N} \Big( |+++\rangle\langle+++| \otimes |j+1\rangle\langle j| \;+\; |---\rangle\langle---| \otimes |j-1\rangle\langle j| \;+\; \alpha \sum_{c\notin\{+++,---\}} |c\rangle\langle c| \otimes |j\rangle\langle j| \Big)$$

or in the form

$$U = \sum_{c\in\{0,1\}^3} |c\rangle\langle c| \otimes D_c,$$

where $|c\rangle = |c_1c_2c_3\rangle$, and the displacement operator $D_c$ is defined as

$$D_c = \begin{cases} \sum_{j=-N}^{N} |j+1\rangle\langle j|, & \text{if } c = 000, \\ \sum_{j=-N}^{N} |j-1\rangle\langle j|, & \text{if } c = 111, \\ \sum_{j=-N}^{N} \alpha\, |j\rangle\langle j| = \alpha\, I_P, & \text{otherwise.} \end{cases}$$

Note: The computational basis states are transformed by $H^{\otimes 3}$ at the beginning of each step before the conditional displacement is applied.